\documentclass{article}

\usepackage[utf8]{inputenc}
\usepackage[T1]{fontenc}

\PassOptionsToPackage{table, dvipsnames}{xcolor}
\usepackage{xcolor}

\PassOptionsToPackage{colorlinks=true, citecolor=blue, linkcolor=blue, urlcolor=black}{hyperref}

\usepackage{arxiv}
\usepackage[utf8]{inputenc}
\usepackage[T1]{fontenc}
\usepackage{url}
\usepackage{booktabs}
\usepackage{nicefrac}
\usepackage{microtype}
\usepackage{lipsum}
\usepackage{graphicx}
\usepackage{natbib}
\usepackage{doi}
\usepackage{algorithm}
\usepackage{multirow}
\usepackage{multicol}
\usepackage{amsmath,amssymb,amsfonts}
\usepackage{amsthm}
\usepackage{mathrsfs}
\usepackage{float}
\usepackage{algorithmic}
\usepackage{bm}
\usepackage{caption}
\usepackage{hyperref}

\title{Adaptive Self-Supervised Surface-Related Multiple Suppression
}

\author{
	{Huan~Song, ~Huanhuan~Tang, ~Wei~Ouyang, ~Weijian~Mao} \\
	Research Center for Computational and Exploration Geophysics, \\State Key Laboratory of Precision Geodesy,\\
    Innovation Academy for Precision Measurement Science and Technology, \\ Chinese Academy of Sciences, Wuhan 430077, China \\
        \And
	\href{https://orcid.org/0000-0001-8868-7967}{\includegraphics[scale=0.06]{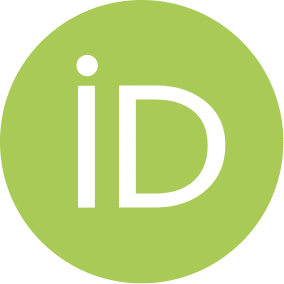}\hspace{1mm}Shijun~Cheng} \\
	Division of Physical Science and Engineering\\
	King Abdullah University of Science and Technology\\
	Thuwal 23955-6900, Saudi Arabia \\
    [3ex]
  $^{*}$Corresponding author: \textbf{Shijun Cheng}~(\texttt{sjcheng.academic@gmail.com})
}

\renewcommand{\shorttitle}{Adaptive Self-Supervised Surface-Related Multiple Suppression}

\hypersetup{
pdftitle={A template for the arxiv style},
pdfsubject={q-bio.NC, q-bio.QM},
pdfauthor={David S.~Hippocampus, Elias D.~Striatum},
pdfkeywords={First keyword, Second keyword, More},
}

\begin{document}
\maketitle

\begin{abstract}
Effective suppression of surface-related multiples is essential to prevent imaging artifacts and erroneous structural interpretations. While conventional approaches rely on accurate priors or subsurface model knowledge, and supervised learning methods require labeled data that are impractical to obtain for real seismic data. To overcome these limitations, a recently proposed self-supervised learning (SSL) framework integrates multi-dimensional convolution (MDC) for multiple generation with a two-stage training strategy, eliminating the need for both prior knowledge and labeled data. However, their approach requires manual selection of a scaling factor to match the amplitudes between the MDC-generated multiples and the true multiples, thus introducing subjectivity and limiting its practical applicability. In this study, we propose an adaptive SSL method that treats the scaling factor as a learnable parameter, jointly optimized with the network weights in a unified single-stage training pipeline. This dynamic scaling implicitly introduces amplitude diversity into the training data, acting as an implicit regularizer that improves the network's robustness to amplitude variations of surface-related multiples. We further design a composite loss function with homoscedastic uncertainty-based adaptive weighting, which automatically balances the contributions of multiple loss terms without manual tuning. Synthetic and field data examples demonstrate that our method robustly and effectively suppresses surface-related multiples while preserving primary reflections, with migration results confirming improved subsurface imaging quality.
\end{abstract}

\section{Introduction}\label{sec:introduction}
Surface-related multiples are seismic events that have undergone at least one downward reflection at the free surface before being recorded. Typically manifesting as repeating events with lower stacking velocity than primaries at equivalent depth. Strong surface-related multiples can entirely obscure weak primary reflections and distort amplitude characteristics, leading to erroneous geological interpretation \citep{verschuur2013seismic}. Therefore, effective suppression of multiples is essential for eliminating structural artifacts and ensuring that seismic images accurately represent subsurface structures for reliable exploration. However, multiples often overlap with primaries in both time and space, making it challenging to subtract them without damaging the primaries.

Over the past several decades, numerous traditional methods have been developed to address the challenge of surface-related multiples suppression, broadly categorized into filtering-based and wave-equation-based approaches. The most common filtering-based methods include predictive deconvolution \citep{porsani2007direct}, Radon transforms (including linear, parabolic, and hyperbolic) \citep{foster1992suppression, jiang2020adaptive, song2022multiple, nowak2006amplitude}, \(f \sim k\) filtering \citep{zhou1994wave}. These methods exploit the differences between multiples and primaries in periodicity, moveout and velocity \citep{hu2008robust}. The main advantages of these methods are their high computational efficiency and their ability to operate without requiring a detailed subsurface velocity model or complex wavefield modeling. However, they rely on underlying assumptions that are often violated in complex geological settings. In areas with dipping layers or complex structures, multiples are no longer periodic. In the Radon or  \(f \sim k\) domain,  the velocity difference between primaries and multiples is too small at deep water or high-velocity areas, where the filter fails to cleanly separate the multiples and may unintentionally remove primary energy.

Wave-equation methods treat multiple reflections as a physical wave propagation problem \citep{hokstad20063d, herrmann2008adaptive, dragoset2010perspective, sui2024seismic}. The well-known wave-equation methods include surface-related multiple elimination (SRME) \citep{verschuur1992adaptive, verschuur1997estimation, berkhout1997estimation}, estimation of primaries by sparse inversion (EPSI) \citep{van2009estimating, van2009estimation, lin2013robust}, closed-loop SRME \citep{lopez2015closed}, and Marchenko theory \citep{he2022elimination}. These methods are entirely data-driven, relying on the free-surface feedback model. The most widely used among them is SRME, which physically reconstructs the multiple model by applying multi-dimensional convolution (MDC) to the recorded seismic data with itself. The predicted model is then refined through adaptive subtraction, where least-squares filters adjust the predicted amplitude and phases to match the raw data, allowing the multiples to be removed while preserving the primaries. Compared to filtering methods, wave-equation methods can handle complex geological settings. However, their computational cost is high, and they require prior knowledge such as accurate wavelet estimation.

To address the limitations of traditional methods, supervised learning (SL) methods have been applied to multiple suppression tasks \citep{siahkoohi2019surface, huan2021application, wang2022seismic, tao2022seismic, liu2022seismic, bao2022surface}. These methods train deep neural networks (NNs) on large paired datasets consisting of noisy seismic records (multiple-contaminated) and corresponding ground-truth labels (multiple-free), through which the network learns a nonlinear mapping that distinguishes multiple features from primaries, acting as an intelligent filter. Compared to wave-equation methods, SL-based approaches offer significantly higher computational efficiency at inference time. However, they depend on high-quality labeled training data, which is impractical to obtain for real seismic records. Relying on labels generated by conventional multiple suppression methods is not only computationally expensive but also limits the network's performance to the accuracy of those preliminary labels. Furthermore, SL-based methods suffer from poor generalization, as a distributional gap exists between synthetic training data and real field data acquired under different conditions \citep{alkhalifah2022mlreal}.

In recent years, several SSL studies have been proposed to overcome the clean-label bottleneck \citep{li2021adaptive, liu2021adaptive, wang2022unsupervised2, wang2022unsupervised1, wang2023surface, wang2023unsupervised, li2023unsupervised, liu2024physics, ma2024u}. These methods begin by using traditional methods to generate predicted multiples. NNs are then employed to map these predictions to the total wavefield, iteratively correcting amplitude and phase distortions between the predicted and true multiples to achieve precise adaptive suppression. However, the quality of multiple suppression in these approaches heavily depends on the accuracy of the predicted multiples provided by the traditional methods, and some require repeated MDC operations, incurring high computational cost. To address these shortcomings, \cite{cheng2025self} developed a flexible SSL framework that leverages MDC to generate multiples directly from the observed data, and employs a two-stage training strategy, which comprises a warm-up and an iterative data refinement stage, to progressively suppress surface-related multiples. This framework eliminates the need for clean labeled data, requires neither wavelets nor subsurface velocity models, and performs MDC only once. Despite these advantages, their method faces a critical practical constraint: each dataset requires a manually selected scaling factor to match the amplitudes between the MDC-generated multiples and the true multiples. As shown in their work, an excessive scaling factor can over-suppress primary reflections, while an insufficient value may leave residual multiples. Although they define a range around an empirically determined median from which the scaling factor is randomly sampled, this procedure introduces subjective user bias and limits the method's applicability to unfamiliar datasets.

In this paper, we propose an adaptive SSL framework built upon the work of \cite{cheng2025self} to resolve its reliance on manual scaling factor selection. Rather than requiring empirical tuning, we treat the scaling factor as a learnable parameter that is jointly optimized with the network weights during training. To facilitate stable and principled optimization, we design a composite loss function comprising a reconstruction loss that captures the spectral and structural features of surface-related multiples, and an $L_1$ regularization term that constrains the update magnitude of the scaling factor, preventing it from collapsing to trivial solutions. Furthermore, we employ homoscedastic uncertainty to adaptively weight the two loss terms, eliminating the need for manual loss balancing and effectively accommodating varying loss scales throughout training. Comparative experiments conducted on the same datasets as \cite{cheng2025self} demonstrate that our method achieves comparable or superior multiple suppression performance using only the warm-up stage, thereby eliminating the need for the subsequent iterative data refinement stage required by Cheng et al.'s method \citep{cheng2025self}. Consequently, our method offers a more robust and streamlined pipeline that preserves the self-supervised advantages of the original work while resolving its key practical limitation.
\section{Review of Self-Supervised Surface-Related Multiple Suppression via MDC}\label{sec:review}
\cite{cheng2025self} proposed an SSL framework for surface-related multiple suppression that operates directly on observed seismic data, requiring neither clean labels, wavelet estimation, nor subsurface velocity models. The framework consists of two components: multiple prediction via MDC and a two-stage training strategy.

In the first component, MDC is applied to the observed data to generate predicted surface-related multiples $d_\text{m}$. The physical basis of MDC follows the free-surface feedback model \citep{wapenaar2011seismic}, in which surface-related multiples can be expressed as a convolution of the recorded wavefield with itself. Although the MDC-generated multiples $d_\text{m}$ may differ from the true multiples in amplitude and phase due to the absence of accurate source wavelets, they preserve the essential spatial and temporal characteristics of surface-related multiples. This property makes them suitable for constructing training pairs within an SSL paradigm.

The second component is a two-stage training strategy comprising a warm-up stage and an iterative data refinement (IDR) stage, both grounded in the Noisier2Noise learning principle \citep{moran2020noisier2noise}. The Noisier2Noise principle states that a NN can learn to denoise a signal by training on pairs where the input is a noisier version of the target, provided that the added noise is statistically independent of the noise already present in the target. Under this condition, minimizing the reconstruction loss between the network output and the noisy target is equivalent, in expectation, to minimizing the loss against the clean signal. In the context of multiple suppression, the surface-related multiples in $d_\text{raw}$ act as the inherent noise, and the MDC-generated multiples $d_\text{m}$ serve as the additional independent noise component injected into the input.

In warm-up stage, the recorded raw seismic data $d_\text{raw}$ serves directly as the pseudo-label. The input data is constructed by augmenting $d_\text{raw}$ with the scaled MDC-predicted multiples:
\begin{equation}\label{eq1}
    d_\text{input}^{(w)} = d_\text{raw} + \alpha d_\text{m},
\end{equation}
where $\alpha$ is a scaling factor that controls the energy of the predicted multiples. The network $f_\theta$ is trained to map $d_\text{input}^{(w)}$ back to $d_\text{raw}$, i.e., to a target that contains fewer multiples than the input. Following the Noisier2Noise principle, this self-supervised objective encourages the network to learn the underlying features of surface-related multiples and suppress them, without requiring any external clean labels.

Although the warm-up stage produces an initial estimate of the multiple-suppressed data, the result may retain residual multiples because $d_\text{raw}$ itself still contains multiples, making it an imperfect pseudo-label. The IDR stage addresses this by progressively improving the quality of the pseudo-labels across training epochs. At each epoch $t$, the pseudo-label is replaced by the network's own prediction from the previous epoch $t-1$:
\begin{equation}\label{eq2}
    d_\text{input}^{(t)} = \hat{d}_\text{pred}^{(t-1)} + \alpha d_\text{m},
\end{equation}
where $\hat{d}_\text{pred}^{(t-1)} = f_\theta(d_\text{input}^{(t-1)})$ is the network output at epoch $t-1$. Since each successive prediction contains progressively fewer multiples than $d_\text{raw}$, the pseudo-label quality improves iteratively. Correspondingly, the input constructed by adding $\alpha d_\text{m}$ to this cleaner pseudo-label becomes a noisier version of a cleaner target, reinforcing the Noisier2Noise condition and enabling the network to refine its suppression capability epoch by epoch.

Despite its advantages, the framework of \cite{cheng2025self} has two critical practical limitations. The first concerns the scaling factor $\alpha$. Since the amplitude of the MDC-generated multiples $d_\text{m}$ does not automatically match that of the true multiples in $d_\text{raw}$, $\alpha$ must be carefully selected to ensure a meaningful training signal. In their implementation, a narrow range is defined around an empirically determined median value, from which $\alpha$ is randomly sampled during training. This median is estimated by visually or quantitatively comparing the amplitude of $\alpha d_\text{m}$ against $d_\text{raw}$, a process that is inherently subjective and dataset-dependent. As demonstrated in their paper, an excessively large $\alpha$ leads to over-suppression of primaries, while an insufficient value results in residual multiples remaining in the output. This manual selection procedure introduces user bias and significantly limits the practical applicability of the framework, particularly for unfamiliar datasets where an appropriate $\alpha$ cannot be determined a prior. The second limitation lies in the computational cost of the IDR stage. Although the iterative refinement of pseudo-labels progressively improves suppression quality, this improvement comes at the expense of a substantially increased training time. Each IDR epoch requires a full forward pass through the network to generate updated pseudo-labels, and the convergence of the refinement process typically demands a large number of iterations. Consequently, the total training time grows significantly compared to the warm-up stage alone, reducing the practical efficiency of the framework for large-scale or time-sensitive applications. Together, these two limitations motivate the development of a more automated and efficient alternative, as proposed in the following section.

\section{Proposed Method}\label{sec:method}
Building upon the SSL framework of \cite{cheng2025self}, we propose an adaptive SSL method that addresses both limitations identified in Section \ref{sec:review}. Our framework retains the MDC-based multiple prediction stage from \cite{cheng2025self} without modification, and focuses on three key innovations: (1) treating the scaling factor $\alpha$ as a learnable parameter to eliminate manual tuning, (2) designing a composite loss function to prevent degenerate optimization, and (3) adopting adaptive loss weighting via homoscedastic uncertainty to automatically balance the two loss terms. As a result, our method enables the warm-up stage alone to achieve robust multiple suppression, rendering the computationally expensive IDR stage unnecessary.

\subsection{Single-Stage Training with Learnable Scaling Factor}

In the framework of \cite{cheng2025self}, the IDR stage is introduced primarily to compensate for the suboptimal suppression achieved during the warm-up stage, which stems from the inaccurate and subjective selection of $\alpha$. We assume that if $\alpha$ is accurately and automatically determined, the warm-up stage alone is sufficient to achieve satisfactory suppression, thereby eliminating the need for iterative refinement.

To this end, we propose treating $\alpha$ as a learnable parameter that is jointly optimized with the network weights $\theta$ via gradient descent throughout training. The input data retains the same construction as in the warm-up stage of \cite{cheng2025self}, i.e., using Equation (\ref{eq1}). The network $f_\theta$ is trained to map $d_\text{input}$ to $d_\text{raw}$. Rather than sampling $\alpha$ from a manually defined range, our framework allows $\alpha$ to evolve autonomously during training, guided by the gradient from the loss function. This dynamic scaling mechanism enables the framework to automatically identify the appropriate amplitude relationship between $d_\text{m}$ and the true multiples in $d_\text{raw}$ for each dataset, without any empirical tuning or visual inspection.

Beyond resolving the subjectivity issue, this learnable $\alpha$ introduces amplitude diversity into the training data as it evolves, acting as an implicit regularizer that improves the network's robustness to amplitude variations of surface-related multiples across diverse datasets. As we will demonstrate in the experiments, this single-stage approach achieves comparable or superior performance to the full two-stage pipeline of \cite{cheng2025self}, while significantly reducing training time and manual effort.

\subsection{Composite Loss Function for Constrained Scaling Optimization}

Treating $\alpha$ as a learnable parameter introduces a critical optimization risk. If the network is trained with only a reconstruction loss, the trivial solution of driving $\alpha$ toward zero becomes the path of least resistance. To see why, note that the network input is $d_\text{input} = d_\text{raw} + \alpha d_\text{m}$, and the training objective is to reconstruct $d_\text{raw}$. The simplest strategy for the network to minimize the reconstruction error is to learn to subtract $\alpha d_\text{m}$ from the input, rather than acquiring any meaningful understanding of the multiple features embedded in $d_\text{m}$. Setting $\alpha \to 0$ eliminates $d_\text{m}$ entirely, causing the network to degenerate into an identity mapping and rendering the self-supervised training ineffective.

To prevent this degenerate solution, we design a composite loss function consisting of two terms. The first is a regression loss $L_\text{reg}$ that enforces consistency between the network prediction and the pseudo-label:
\begin{equation}\label{eq3}
    L_\text{reg} = \|d_\text{raw} - d_\text{pred}\|_1.
\end{equation}
The second is a constraint loss $L_\text{cons}$ that anchors $\alpha$ toward a prescribed target value, preventing it from collapsing to zero:
\begin{equation}\label{eq4}
    L_\text{cons} = \|\alpha - 0.5\|_1.
\end{equation}
The target value of 0.5 is chosen to ensure that the predicted multiples $\alpha d_\text{m}$ retain a meaningful energy level relative to $d_\text{raw}$ throughout training. The initial value of $\alpha$ is set to a small non-zero value (e.g., $\alpha_0 = 0.01$). As training progresses, $L_\text{cons}$ gradually steers $\alpha$ toward 0.5, while $L_\text{reg}$ simultaneously drives the network to learn the underlying features of surface-related multiples. Therefore, the two losses encourage $\alpha$ to settle at a meaningful non-trivial value, ensuring that the predicted multiples remain an informative and non-degenerate component of the training input throughout optimization.

A standard formulation of the total loss with fixed weights is:
\begin{equation}\label{eq5}
    L_\text{total}(\lambda_1, \lambda_2) = \lambda_1 L_\text{reg} + \lambda_2 L_\text{cons},
\end{equation}
where $\lambda_1$ and $\lambda_2$ are manually specified hyperparameters. However, determining appropriate values for these weights is non-trivial: the two loss terms may operate at substantially different scales, and their relative magnitudes can shift dynamically as $\alpha$ and $\theta$ evolve during training. This motivates the need for an adaptive weighting strategy, as described in the following subsection.

\begin{figure*}[!t]
    \centering
    \includegraphics[width=1\linewidth]{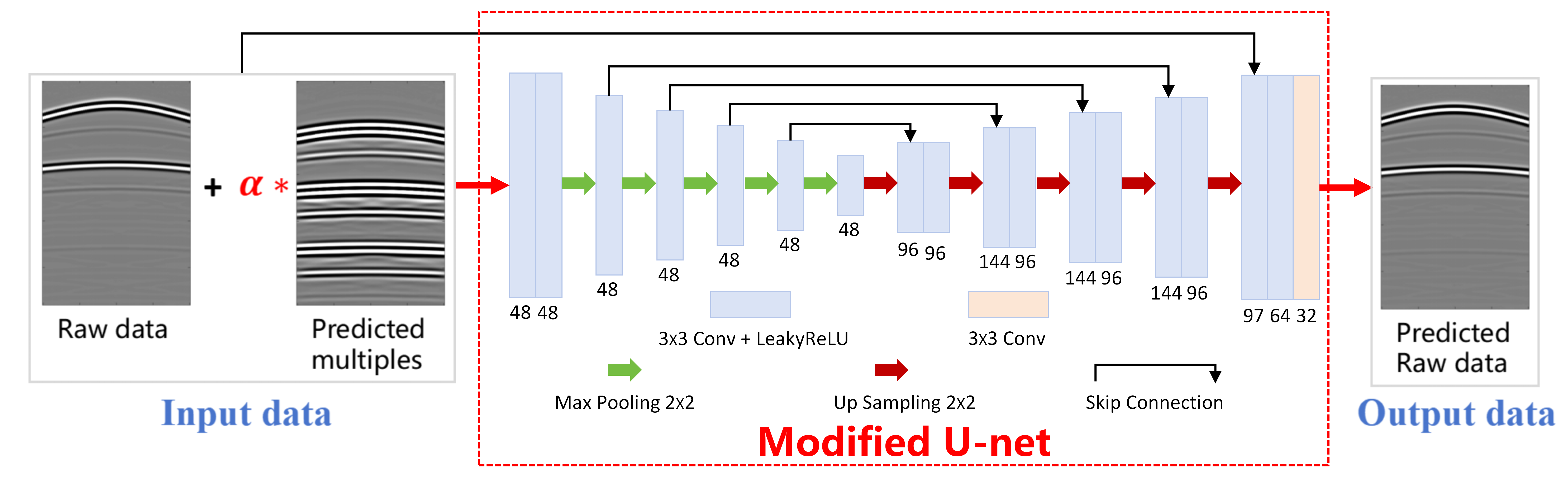}
    \caption{The SSL network structure used in our study. The input data is the sum of raw data and $\alpha$ times of MDC-based predicted multiples. $\alpha$ is a scaling factor that changes the energy of the predicted multiples. The output is the predicted raw data. The network is a modified U-Net, contains five max pooing processes (green arrows) in the encoder and five up sampling processes (dark-red arrows) in the decoder, with skip connections (black arrows) bridging the corresponding layers.}
    \label{fig1}
\end{figure*}

\subsection{Adaptive Loss Weighting via Homoscedastic Uncertainty}
To eliminate the need for manual loss weight selection, we adopt the homoscedastic uncertainty framework proposed by Kendall et al. \citep{Kendall_2018} for multi-task learning. In this formulation, each loss term is associated with a learnable uncertainty parameter that automatically modulates its contribution to the total loss. The adaptively weighted total loss is defined as:
\begin{equation}\label{eq6}
    L_\text{total}(\sigma_1, \sigma_2) = \frac{1}{2\sigma_1^2} L_\text{reg} + 
    \frac{1}{2\sigma_2^2} L_\text{cons} + \log\sigma_1 + \log\sigma_2,
\end{equation}
where $\sigma_1$ and $\sigma_2$ are learnable parameters representing the homoscedastic uncertainty of $L_\text{reg}$ and $L_\text{cons}$, respectively. The uncertainty-based weights $\frac{1}{2\sigma_1^2}$ and $\frac{1}{2\sigma_2^2}$ decrease as the corresponding $\sigma$ increases, automatically down-weighting loss terms with higher uncertainty. The logarithmic penalty terms $\log\sigma_1$ and $\log\sigma_2$ serve as regularizers that prevent the model from trivially minimizing the total loss by driving $\sigma_1$ and $\sigma_2$ to infinity.

This formulation offers two key advantages over the fixed-weight counterpart in Eq. \eqref{eq5}. First, since $\sigma_1$ and $\sigma_2$ are optimized jointly with $\theta$ and $\alpha$ via back-propagation, the loss weighting is entirely self-tuning, eliminating the need for manual hyperparameter search. Second, the framework naturally accommodates the varying loss scales that arise during training: if $L_\text{total}$ becomes disproportionately large relative to $L_\text{total}$, the model automatically increases $\sigma_1$ to reduce its relative contribution, maintaining a stable and balanced optimization landscape throughout training. The complete set of trainable parameters in our framework is therefore $\{\theta, \alpha, \sigma_1, \sigma_2\}$, all optimized jointly through a single back-propagation pass at each training iteration.

\subsection{Network Architecture}
The network architecture used in this paper is shown in Figure \ref{fig1}, which follows the same design as that adopted in \cite{cheng2025self} to ensure a fair comparison.  We employ a modified U-Net with an encoder-decoder structure and skip connections to facilitate multi-scale feature fusion. In the encoder path, five successive $2\times2$ max-pooling operations progressively reduce the spatial resolution while preserving critical seismic features, with the number of feature channels doubled at each stage. Each encoding and decoding block consists of a $3\times3$ convolution followed by a LeakyReLU activation. In the decoder path, five $2\times2$ up-sampling operations symmetrically reconstruct the spatial dimensions, with skip connections bridging the corresponding encoder and decoder layers to reintroduce high-resolution details that aid the network in distinguishing multiples from primaries. A final $3\times3$ convolution collapses the feature maps into a single-channel output to produce the predicted data $d_\text{pred}$.
\section{Synthetic examples}\label{sec:syn_examples}
We first evaluate the proposed method on two synthetic datasets derived from a layered model and an Otway model, respectively, both of which are identical to those used in \cite{cheng2025self}, hereafter referred to as SSL-MDC. SSL-MDC serves as the benchmark for all comparisons throughout the synthetic and field data experiments. Unless otherwise stated, all training hyperparameters follow those of SSL-MDC, and any deviations are explicitly noted.

\begin{figure}[!t]
    \centering
    \includegraphics[width=0.7\linewidth]{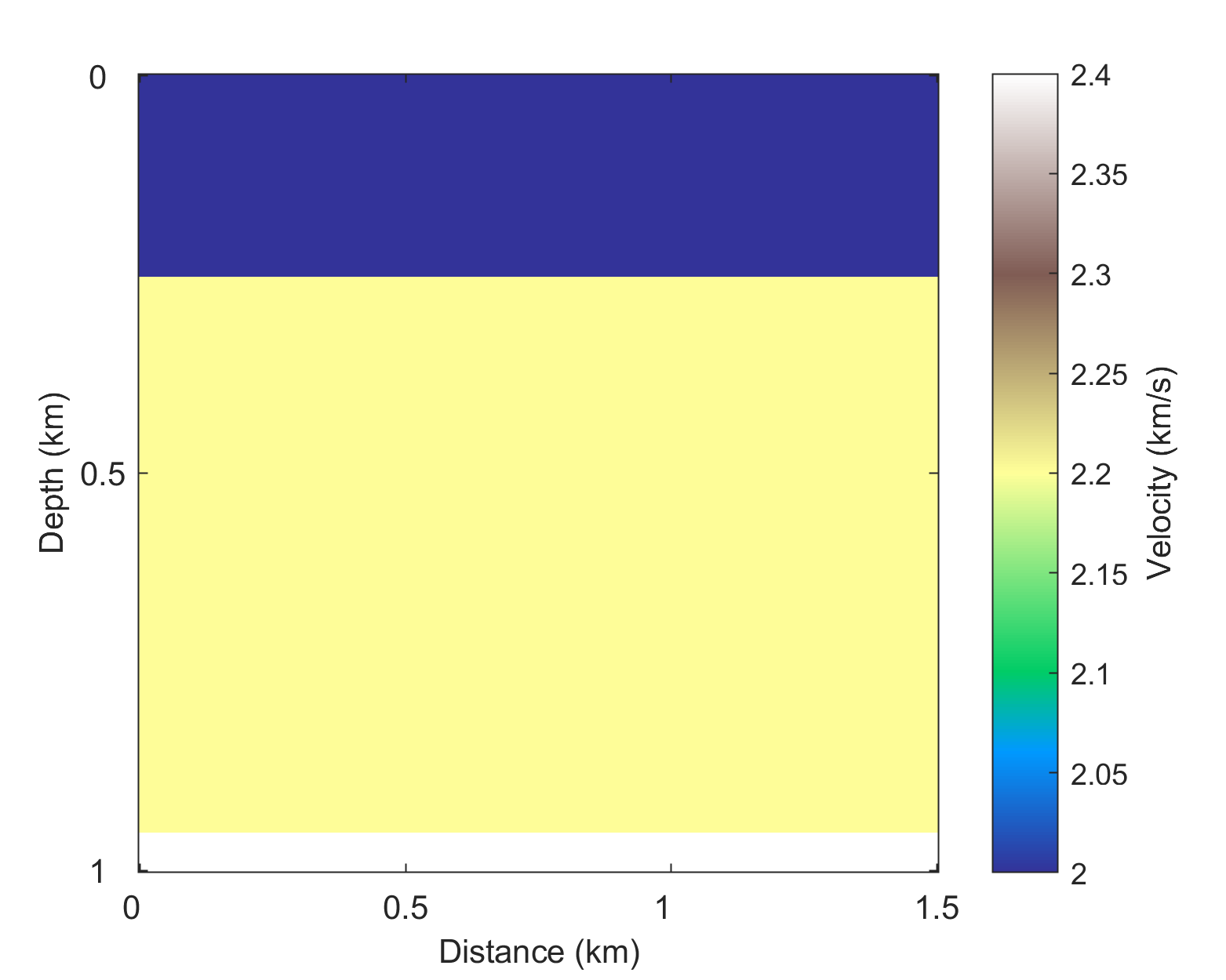}
    \caption{The layered velocity model.}
    \label{fig2}
\end{figure}

\begin{figure*}[!t]
    \centering
    \includegraphics[width=1\linewidth]{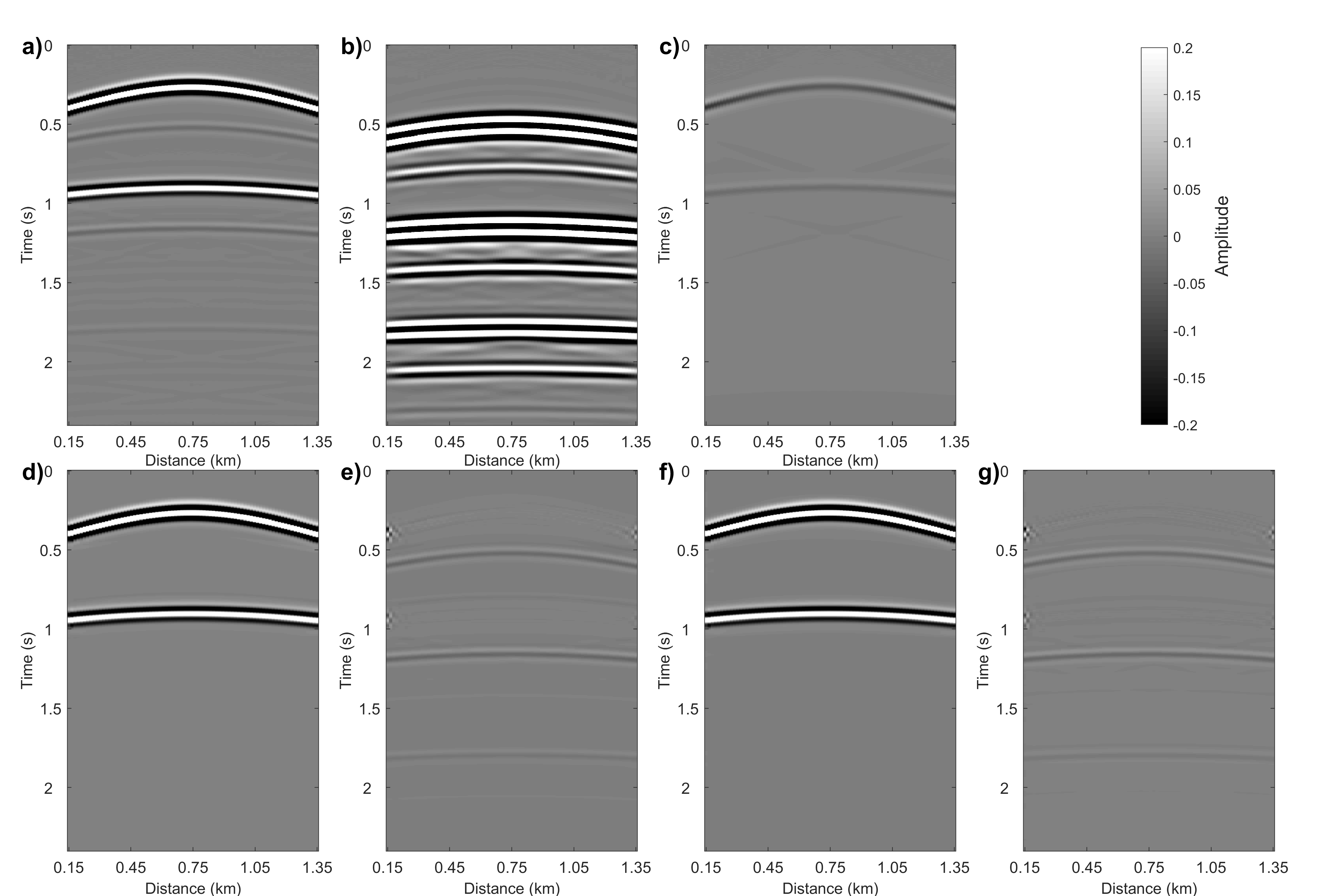}
    \caption{Multiple suppression results for the layered model. (a) A representative shot gather from the layered model with a free-surface boundary. (b) Predicted multiples generated by using the MDC operation to the raw shot gather in (a). (c) A reference shot gather simulated with an absorbing boundary. (d) Multiple suppression result obtained using our method. (e) Residual between the raw data (a) and our multiple suppression result (d). (f) Multiple suppression result obtained by SSL-MDC. (g) Residual between the raw data (a) and the multiple suppression result (f).}
    \label{fig3}
\end{figure*}

\subsection{Layer model}
The layered model, which is shown in Figure \ref{fig2}, consists of three horizontal layers with thicknesses of 0.25 km, 0.7 km, and 0.05 km from top to bottom, and their corresponding P-wave velocities are 2.0 km/s, 2.2 km/s, and 2.4 km/s, respectively. Figure \ref{fig3}a shows a representative shot gather synthesized from the layered model, which contains both primaries and surface-related multiples. The predicted multiples generated by the MDC operator are shown in Figure \ref{fig3}b. The MDC-predicted multiples share the same spatial and temporal locations as the true surface-related multiples in Figure \ref{fig3}a, but differ in amplitude and wavelength, which is consistent with the expected behavior of MDC under the absence of accurate source wavelets. For reference, Figure \ref{fig3}c shows a shot gather simulated with absorbing boundary conditions, which contains only primaries. Although the amplitudes of the primaries in Figure \ref{fig3}c differ from those in Figure \ref{fig3}a due to the different boundary conditions, their spatial locations and phases remain consistent, making Figure \ref{fig3}c a reliable reference for identifying primaries in the suppression results.

In terms of training configuration, the batch size (32), learning rate (2e-4), and optimizer (AdamW) are kept identical to those of SSL-MDC. The total number of training epochs is set to 50, compared to 160 in SSL-MDC, reflecting the elimination of the computationally expensive IDR stage in our framework.

The raw data in Figure \ref{fig3}a is fed into the trained network, and the suppression result is shown in Figure \ref{fig3}d. Figure \ref{fig3}e presents the corresponding residual, obtained by subtracting Figure \ref{fig3}d from Figure \ref{fig3}a. The results demonstrate that our method effectively suppresses surface-related multiples while preserving the primaries. For comparison, Figures \ref{fig3}f and \ref{fig3}g display the suppression result and residual obtained by SSL-MDC. The two sets of results are qualitatively comparable, confirming that treating the scaling factor as a learnable parameter is a viable alternative to manual selection, and that the proposed composite loss function with adaptive weighting effectively guides the optimization. Minor amplitude leakages near the boundaries are observable in both residuals, which are attributable to the checkerboard effect inherent to the U-Net architecture and are not expected to affect subsequent imaging.

\begin{figure}[!t]
    \centering
    \includegraphics[width=0.7\linewidth]{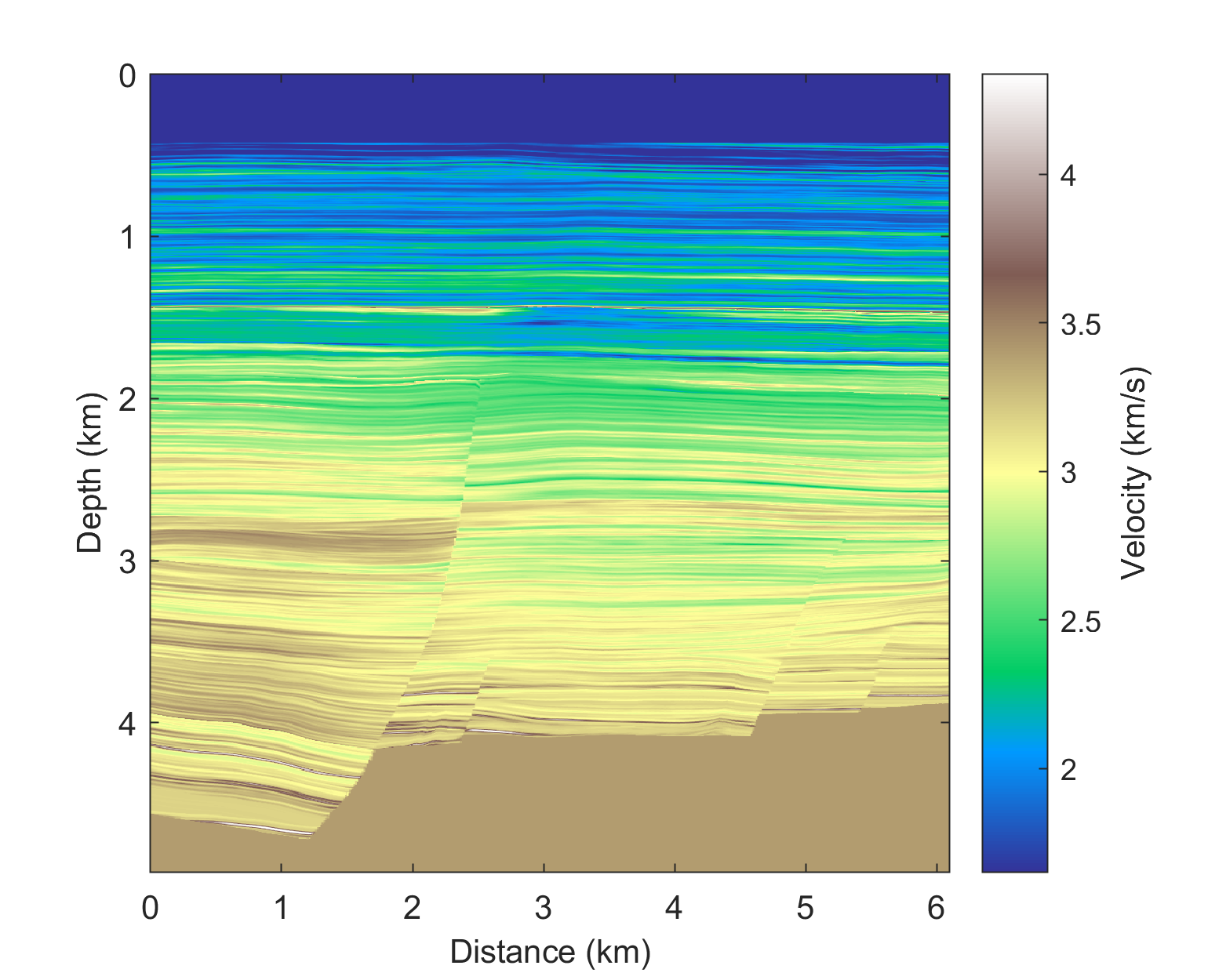}
    \caption{The Otway velocity model.}
    \label{fig4}
\end{figure}

\begin{figure*}[!t]
    \centering
    \includegraphics[width=1\linewidth]{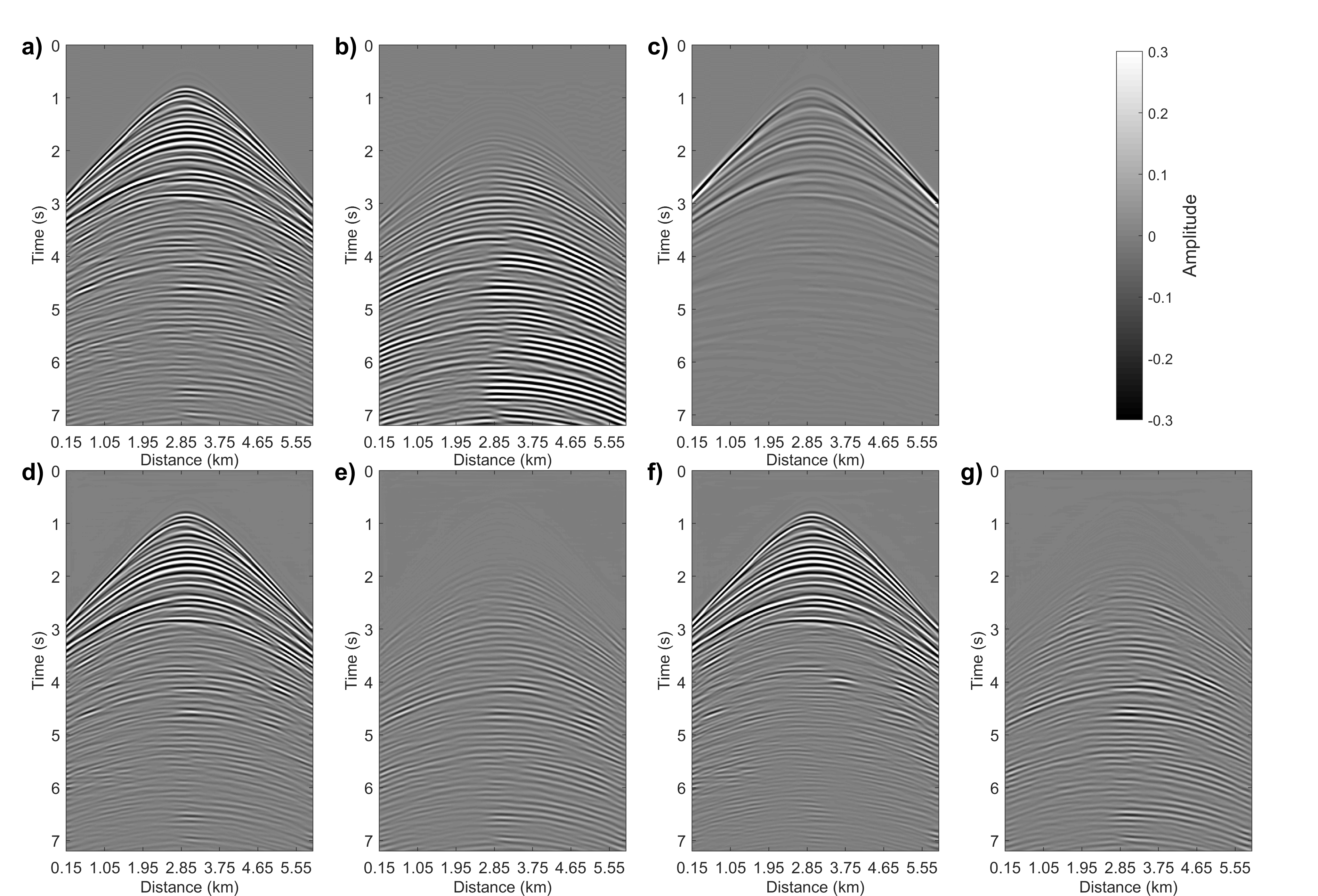}
    \caption{Multiple suppression results for the Otway model. (a) A representative shot gather synthesized with a free-surface boundary. (b) Predicted surface-related multiples generated by applying the MDC operator to the raw shot gather in (a). (c) A reference shot gather simulated with an absorbing boundary condition. (d) Multiple suppression result obtained by the proposed method. (e) Residual between the raw data in (a) and the suppression result in (d). (f) Multiple suppression result obtained by SSL-MDC. (g) Residual between the raw data in (a) and the suppression result in (f).}
    \label{fig5}
\end{figure*}

\begin{figure*}[!t]
    \centering
    \includegraphics[width=1\linewidth]{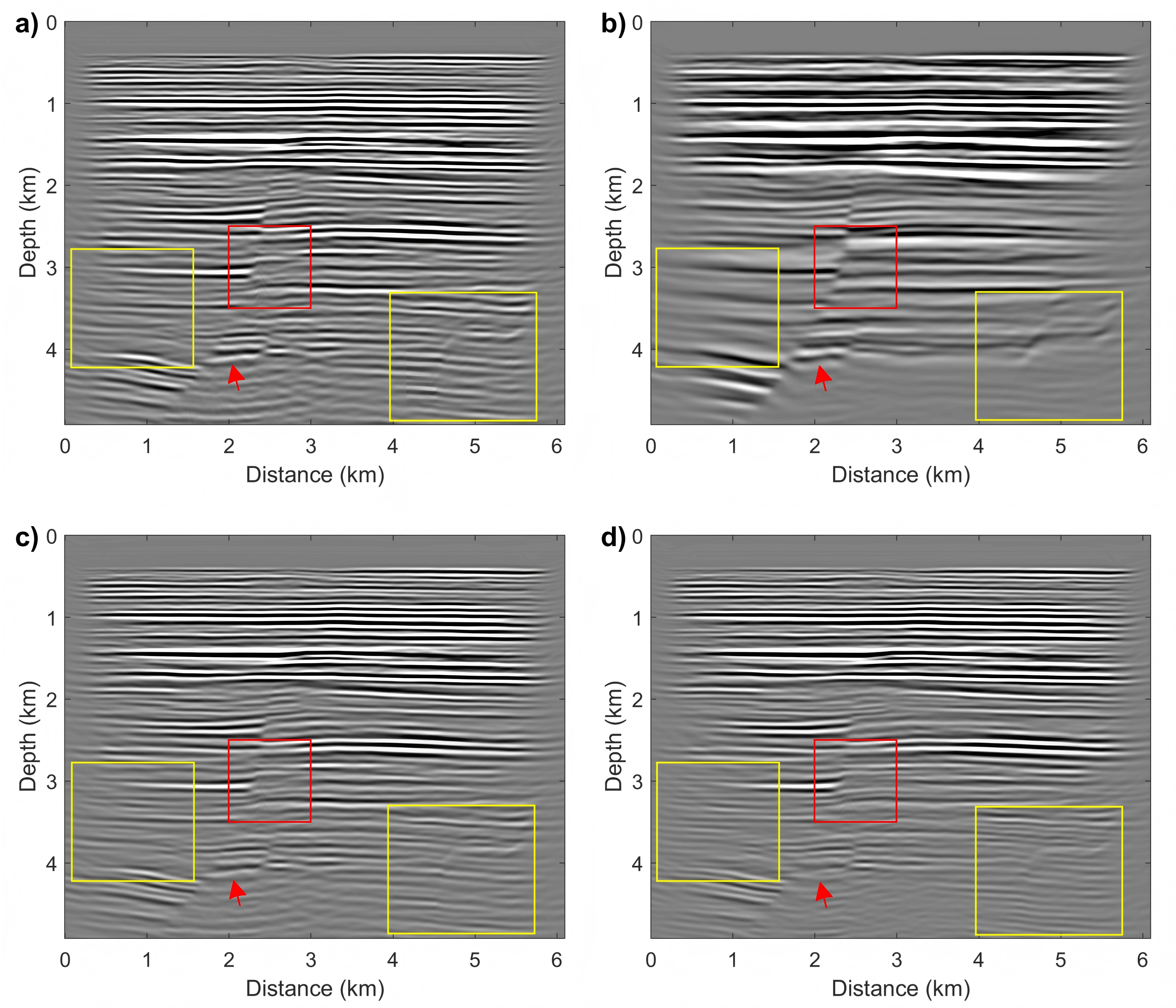}
    \caption{Migrated images for the Otway model. (a) Migration of the raw data containing surface-related multiples. (b) Migration of the reference data simulated with an absorbing boundary condition. (c) Migration of the multiple suppression result obtained by the proposed method. (d) Migration of the multiple suppression result obtained by SSL-MDC. The red rectangles highlight regions where surface-related multiples are effectively suppressed. The red arrow marks a legitimate geological layer that is clearly visible in (a), (b), and (c), but is largely absent in (d), indicating over-suppression by SSL-MDC. The yellow rectangles highlight regions where the proposed method maintains high structural consistency with the raw and reference data, whereas (d) exhibits altered layer geometries and newly introduced artifacts.} 
    \label{fig6}
\end{figure*}

\begin{figure*}[!t]
    \centering
    \includegraphics[width=1\linewidth]{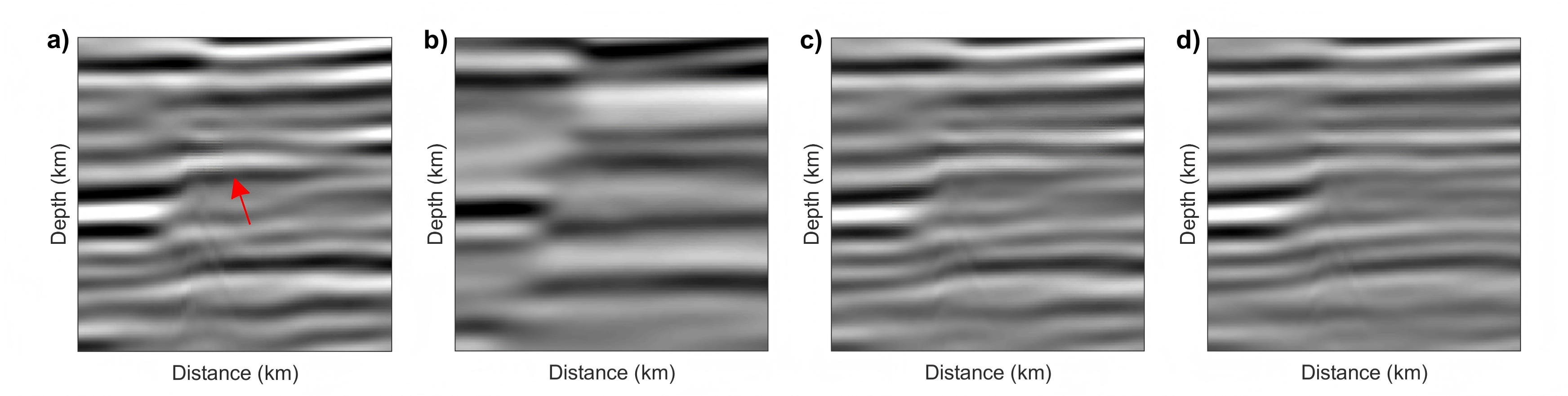}
    \caption{Zoomed-in views of the red rectangular regions marked in Figure \ref{fig6}. The red arrow indicates a spurious reflector introduced by surface-related multiples in the raw data migration, which is successfully removed by both the proposed method and SSL-MDC.}. 
    \label{fig7}
\end{figure*}

\begin{figure*}[!t]
    \centering
    \includegraphics[width=1\linewidth]{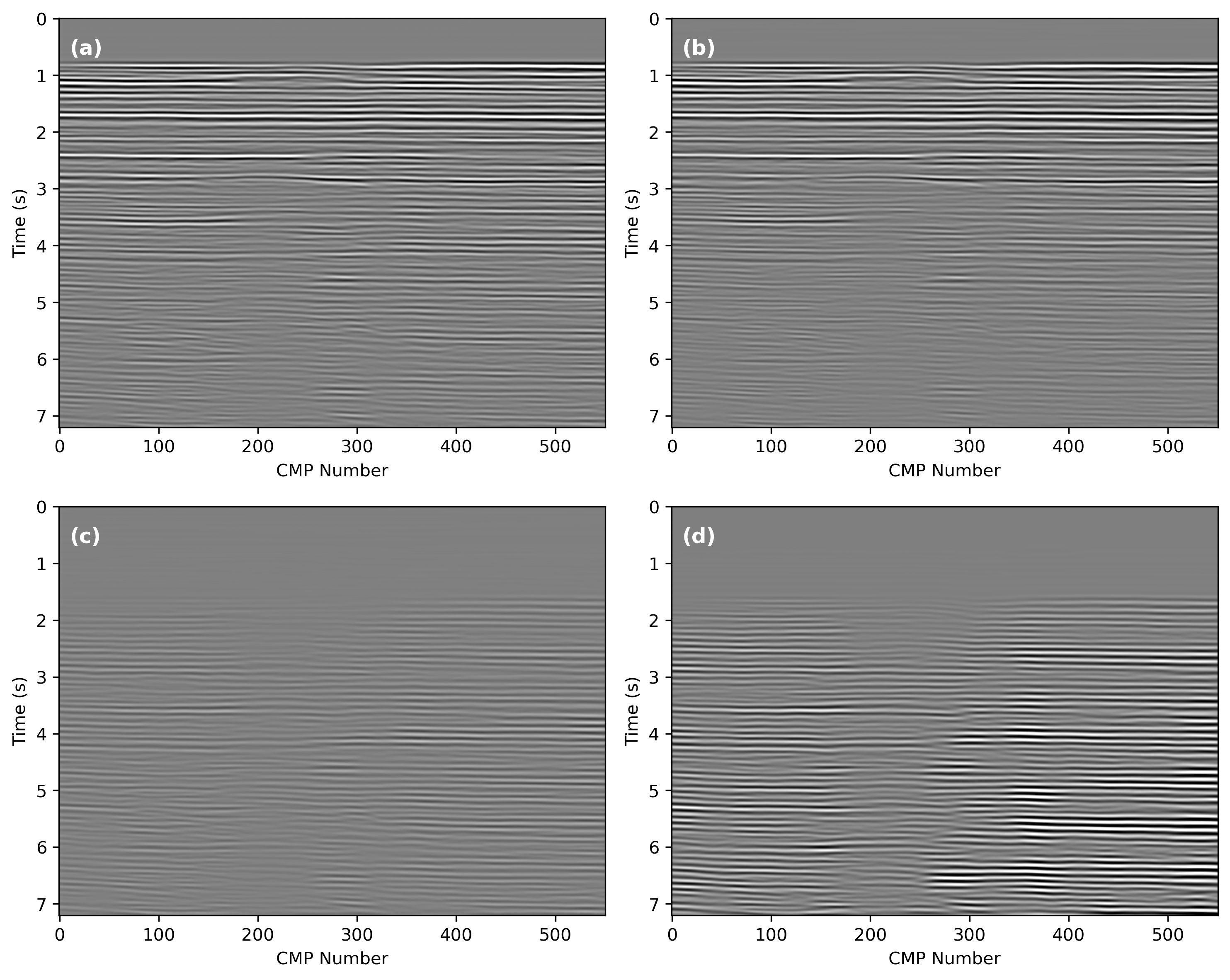}
    \caption{Nearest-offset stacking profiles for the Otway model. (a) Stacking profile of the raw data containing surface-related multiples. (b) Stacking profile of the multiple suppression result obtained by the proposed method. (c) Isolated multiples extracted as the difference between (a) and (b). (d) Reference multiples predicted via the MDC operator.} 
    \label{fig8}
\end{figure*}

\begin{figure*}[!t]
    \centering
    \includegraphics[width=1\linewidth]{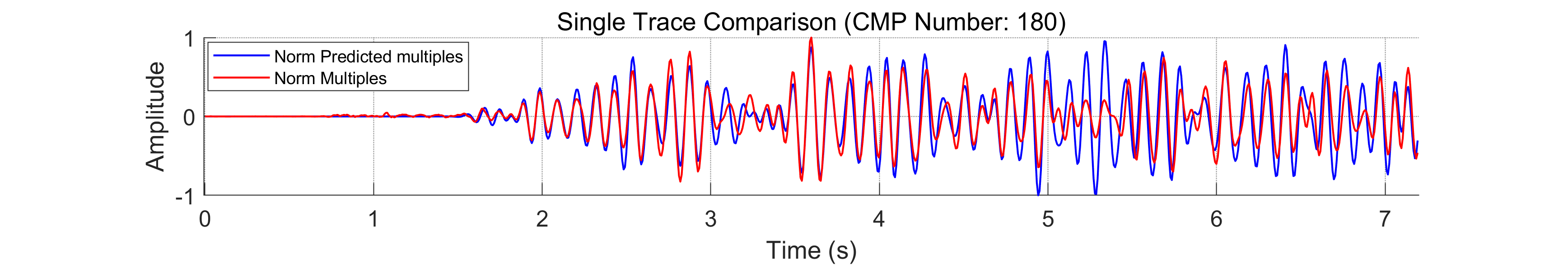}
    \caption{Waveform comparison of a single CMP trace (CMP 180) extracted from the stacking profiles in Figures \ref{fig8}c and \ref{fig8}d. The red curve represents the normalized multiples extracted by the proposed method, and the blue curve represents the normalized reference multiples predicted via the MDC operator. Both traces are normalized by their respective maximum absolute amplitudes to facilitate waveform comparison.}
    \label{fig9}
\end{figure*}

\subsection{Otway Model}
To further evaluate the performance of the proposed method on a more geologically complex scenario, we consider the Otway model, which contains numerous thin layers and several faults, as shown in Figure \ref{fig4}. Figure \ref{fig5}a shows a representative shot gather synthesized from the Otway model, and Figure \ref{fig5}b shows the corresponding predicted surface-related multiples generated by the MDC operation. Consistent with the layered model results, the MDC-generated multiples share the same spatial and temporal positions as the true surface-related multiples in Figure \ref{fig5}a, but exhibit differences in amplitude and wavelength. For reference, Figure \ref{fig5}c shows a shot gather simulated with absorbing boundary conditions. In terms of training configuration, the total number of training epochs is set to 50, compared to 80 in SSL-MDC, while all other hyperparameters remain unchanged.

Figure \ref{fig5}d presents our multiple suppression result using the trained network, with the corresponding residual shown in Figure \ref{fig5}e. Comparing Figure \ref{fig5}e with Figure \ref{fig5}b, the spatial and temporal positions of the suppressed energy closely coincide with those of the MDC-generated multiples, confirming that our method effectively suppresses surface-related multiples without significantly damaging the primaries. For comparison, Figures \ref{fig5}f and \ref{fig5}g display the suppression result and residual obtained by SSL-MDC. A direct comparison between Figures \ref{fig5}d and \ref{fig5}f reveals a notable difference: in the SSL-MDC result, primary energy is visibly damaged between 3.2 s and 7.0 s, and primary events appear discontinuous between 1.5 km and 5.0 km. In contrast, our result preserves the continuity and amplitude of the primaries throughout, demonstrating the advantage of the learnable scaling factor in preventing over-suppression.

To further validate the suppression quality, we compare the migrated images obtained from: the raw data containing surface-related multiples (Figure \ref{fig6}a), the reference data simulated with absorbing boundary conditions (Figure \ref{fig6}b), our suppression result (Figure \ref{fig6}c), and the SSL-MDC suppression result (Figure \ref{fig6}d). To better reveal the impact of surface-related multiples on the migration, the regions marked by red rectangles in Figure \ref{fig6} are enlarged in Figure \ref{fig7}. As indicated by the red arrow in Figure \ref{fig7}a, the multiples introduce a spurious reflector in the raw data migration. Both our method and SSL-MDC successfully suppress this artifact, as confirmed by the clean images in Figures \ref{fig7}c and \ref{fig7}d.

Despite this shared success, notable differences emerge at a finer scale. As highlighted by the red arrow in Figures \ref{fig6}a, \ref{fig6}b, and \ref{fig6}c, a legitimate geological layer is consistently visible across the raw data, reference, and our result. However, this layer is largely absent in Figure \ref{fig6}d, indicating that SSL-MDC suffers from over-suppression at this location. Furthermore, within the yellow rectangular regions in Figure \ref{fig6}, our result maintains high structural consistency with both the raw data and the reference image, whereas the SSL-MDC result exhibits altered layer geometries and the introduction of new artifacts. These observations collectively demonstrate that our adaptive framework achieves a superior balance between multiple suppression and the preservation of true seismic structures.

For additional quantitative validation, we examine the nearest offset stacking profiles. Figure \ref{fig8}a presents the stacking profile of the raw data, and Figure \ref{fig8}b shows the corresponding profile after our multiple suppression. The difference between these two profiles, which represents the suppressed multiples extracted by our method, is shown in Figure \ref{fig8}c. Figure \ref{fig8}d provides the MDC-generated multiples as a reference. A comparison between Figures \ref{fig8}c and \ref{fig8}d reveals that the suppressed energy shares nearly identical spatial and temporal characteristics with the MDC-generated multiples, confirming that the suppressed energy corresponds accurately to surface-related multiples rather than primary reflections.

To perform a more rigorous waveform-level analysis, we extract a single CMP trace (CMP number 180) from both Figures \ref{fig8}c and \ref{fig8}d and compare them in Figure \ref{fig9}, with both traces normalized by their respective maximum absolute amplitudes. The two traces exhibit a high degree of waveform consistency, with peaks and troughs showing excellent temporal alignment throughout the entire time range. This trace-level agreement provides further confirmation that the suppressed energy accurately captures the characteristics of surface-related multiples, validating the effectiveness of the proposed method.

\begin{figure*}[!t]
    \centering
    \includegraphics[width=1\linewidth]{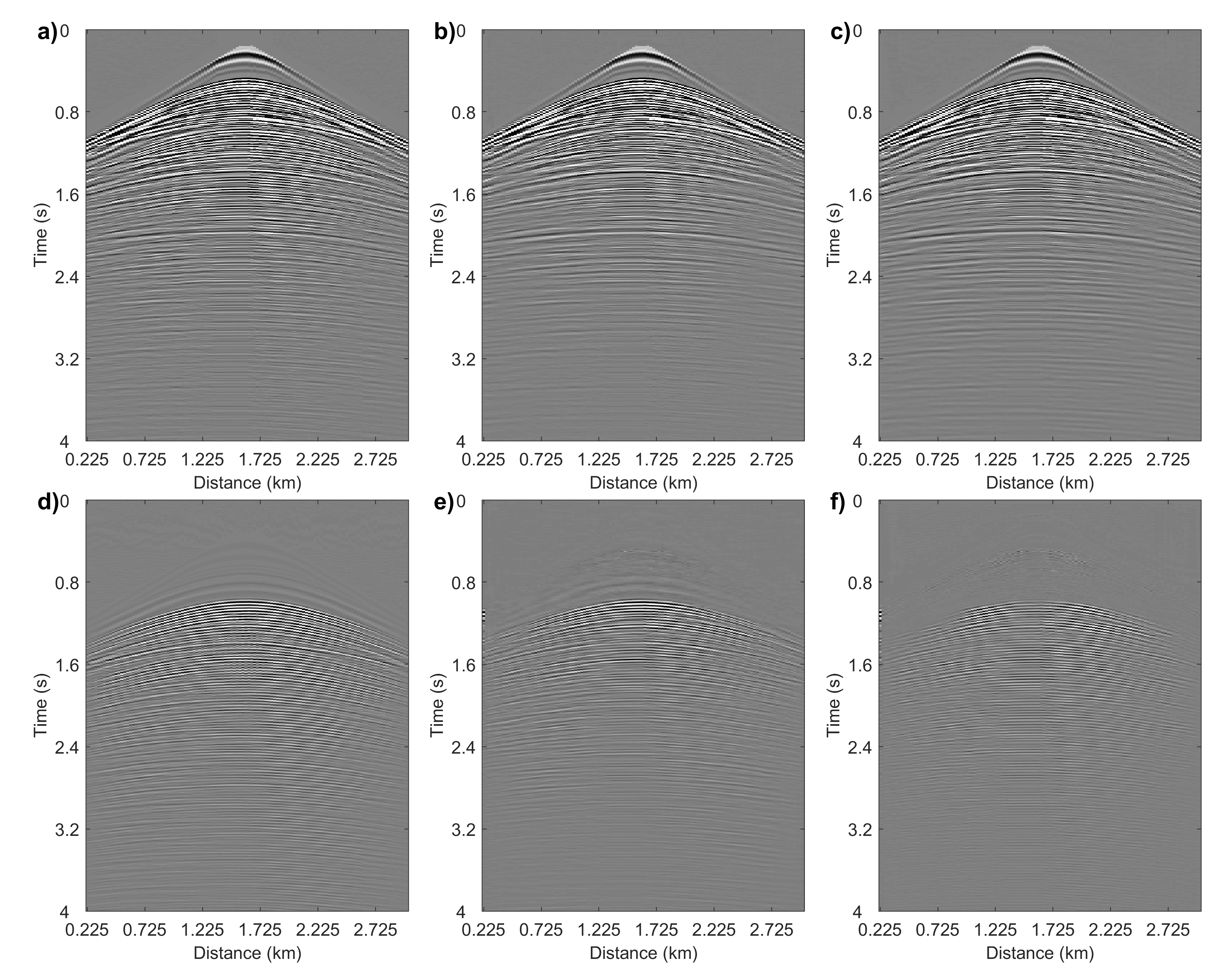}
    \caption{Multiple suppression results for the field data. (a) A representative preprocessed shot gather by \cite{cheng2025self}. (b) Multiple suppression result obtained by the proposed method. (c) Multiple suppression result obtained by SSL-MDC. (d) Reference surface-related multiples  predicted via the MDC operator, displayed with scaled amplitudes for enhanced visual comparison. (e) Multiples extracted by the proposed method, representing the difference between (a) and (b). (f) Multiples extracted by SSL-MDC, representing the difference between (a) and (c).}
    \label{fig10}
\end{figure*}

\begin{figure*}[!t]
    \centering
    \includegraphics[width=1\linewidth]{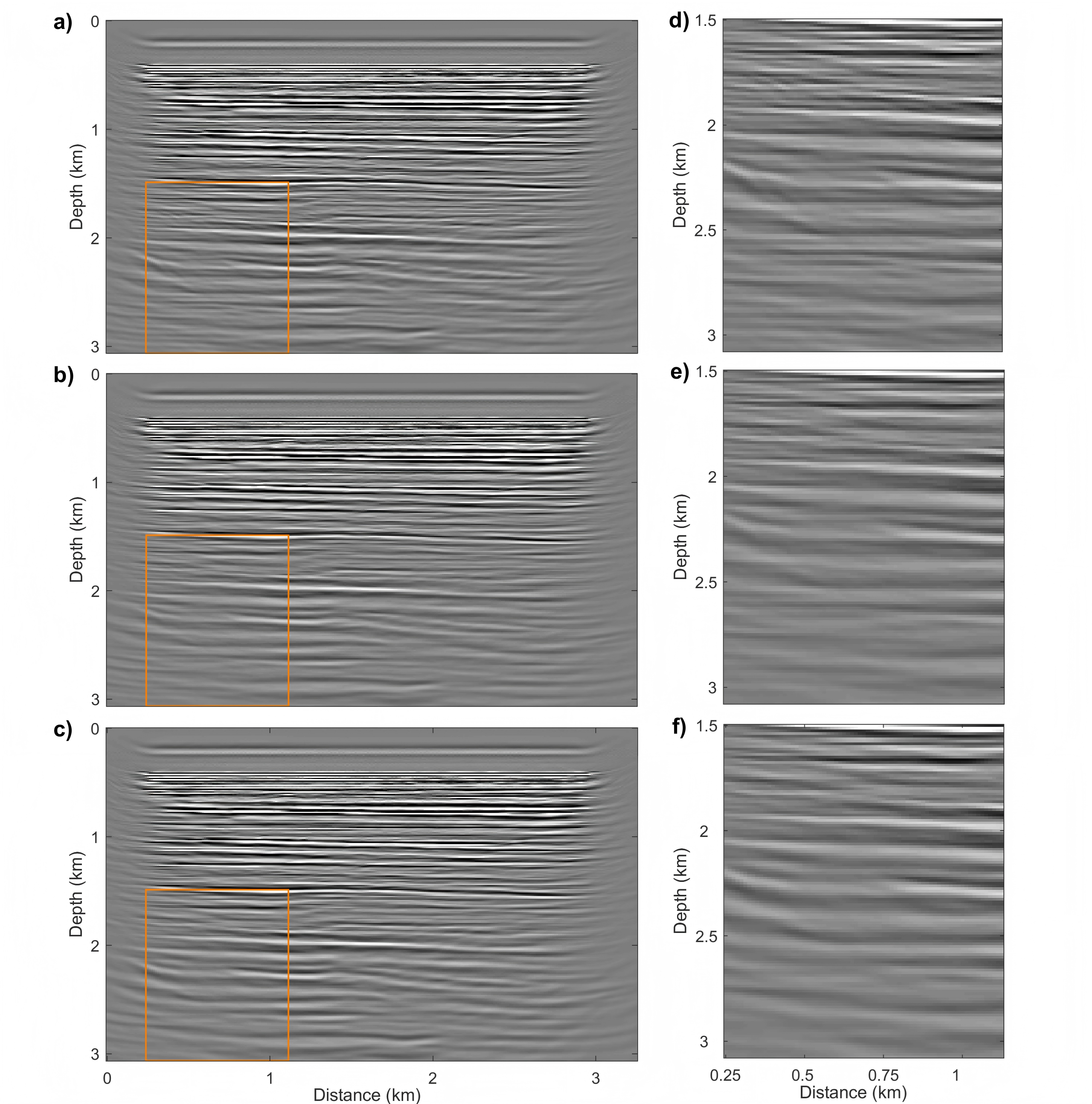}
    \caption{Migrated images for the field data. (a) Migration of the raw data containing surface-related multiples. (b) Migration of the multiple suppression result obtained by the proposed method. (c) Migration of the multiple suppression result obtained by SSL-MDC. (d-f) Zoomed-in views of the yellow rectangular regions highlighted in (a), (b), and (c), respectively. }
    \label{fig11}
\end{figure*}

\section{Field data application}\label{sec:field_examples}
Following the validation of our method on two synthetic models, we further evaluate the proposed method on field data. The dataset used here is identical to that in SSL-MDC, obtained from the publicly available Mobil Viking Graben Line 12 marine AVO dataset. This dataset was acquired using a marine streamer configuration, which introduces two practical challenges for MDC-based multiple prediction. First, near-offset traces are missing due to the physical separation between the source and the near end of the streamer, violating the assumption of full-offset coverage required by MDC. Second, streamer acquisition records only one-sided wavefields, whereas MDC requires source-over-receiver data. To address these issues, SSL-MDC applied a series of preprocessing steps to the raw dataset prior to MDC: near-offset traces were reconstructed to compensate for the missing near-offset coverage, and shot-receiver reciprocity was then applied to transform the one-sided streamer data into the source-over-receiver format required by MDC. In this study, we directly adopt the preprocessed data provided by SSL-MDC, ensuring that any differences in suppression performance between the two methods are solely attributable to the proposed adaptive scaling and loss function designs, rather than to differences in data preparation. A representative shot gather from the prepocessed dataset is shown in Figure \ref{fig10}a. The total number of training epochs is set to 50, while all other hyperparameters are kept identical to those of SSL-MDC to ensure a fair comparison.

Figures \ref{fig10}b and \ref{fig10}c present the multiple suppression results obtained by the proposed method and SSL-MDC, respectively. To facilitate verification of the suppressed multiples, the MDC operator is applied to generate predicted multiples, displayed with a scaled amplitude range for enhanced visualization in Figure \ref{fig10}d. The multiples extracted by the proposed method and SSL-MDC are shown in Figures \ref{fig10}e and \ref{fig10}f, respectively. Both methods suppress significant surface-related multiples across the gather. However, the proposed method demonstrates superior suppression performance in the far-offset regions and deeper time zones, where multiple contamination is typically more difficult to address. These results confirm the effectiveness of the proposed approach in complex field environments.

For further validation, we perform migration on the raw field data (Figure \ref{fig11}a), the suppression result of the proposed method (Figure \ref{fig11}b), and that of SSL-MDC (Figure \ref{fig11}c). To better illustrate the impact of surface-related multiples on imaging quality, the regions marked by yellow rectangles in the left panels of Figure \ref{fig11} are shown as zoomed-in views in the right panels (Figures \ref{fig11}d, \ref{fig11}e, and \ref{fig11}f). In the zoomed-in view of the raw data migration (Figure \ref{fig11}d), the deeper region is dominated by a large number of high-frequency, fine-layered events. These events are indicative of surface-related multiples, which accumulate additional travel time with each surface reflection and therefore manifest as densely packed apparent reflectors at greater depths in the migrated image. After multiple suppression, these spurious fine-layered events are largely removed in both the proposed method result (Figure \ref{fig11}e) and the SSL-MDC result (Figure \ref{fig11}f), yielding a cleaner and more geologically interpretable image of the subsurface structure.

\begin{figure*}[!t]
    \centering
    \includegraphics[width=1\linewidth]{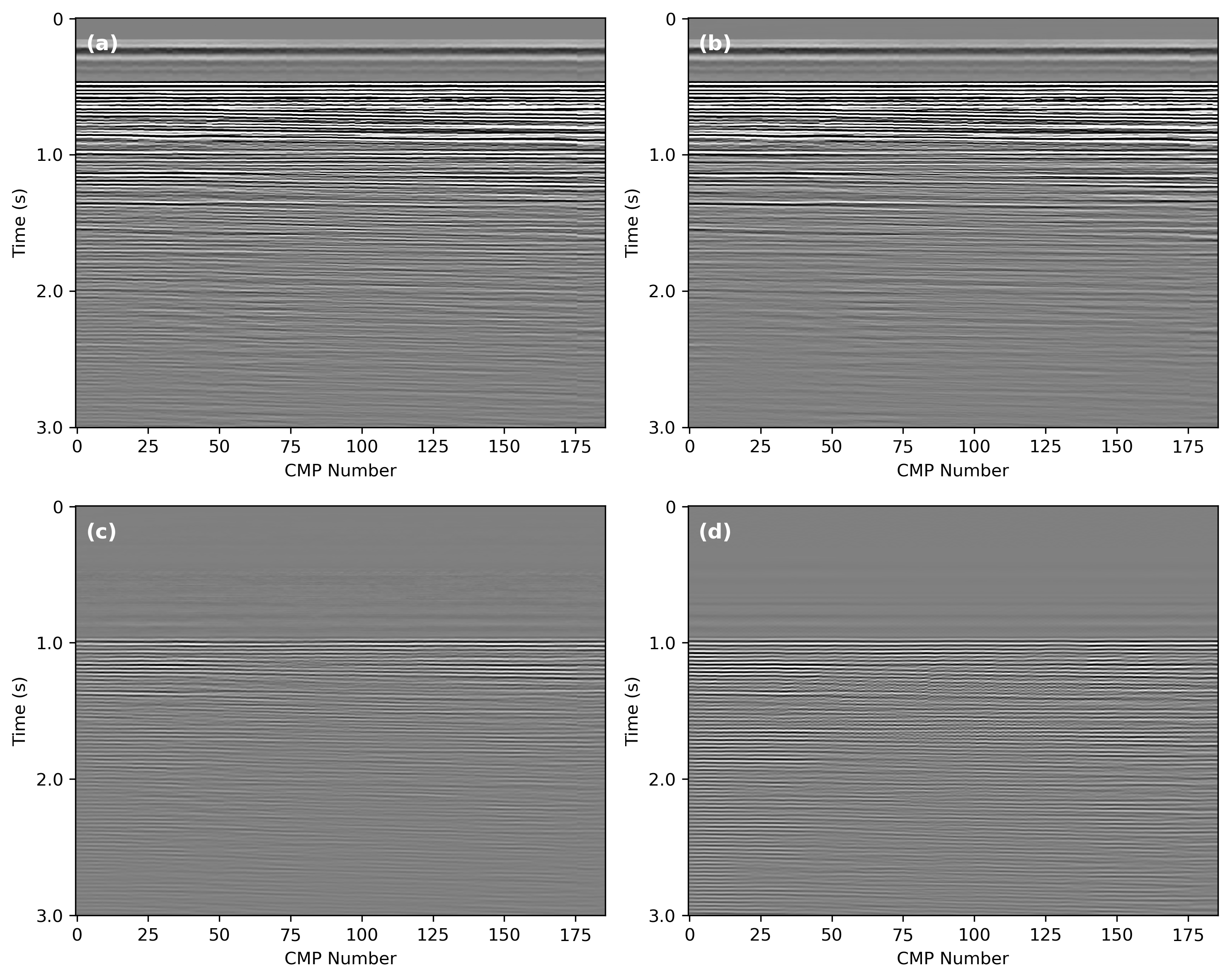}
    \caption{Nearest-offset stacking profiles for the field data. (a) Stacking profile of the raw data containing surface-related multiples. (b) Stacking profile of the multiple suppression result obtained by the proposed method. (c) Isolated multiples extracted as the difference between (a) and (b). (d) Reference multiples predicted via the MDC operation.}
    \label{fig12}
\end{figure*}

\begin{figure*}[!t]
    \centering
    \includegraphics[width=1\linewidth]{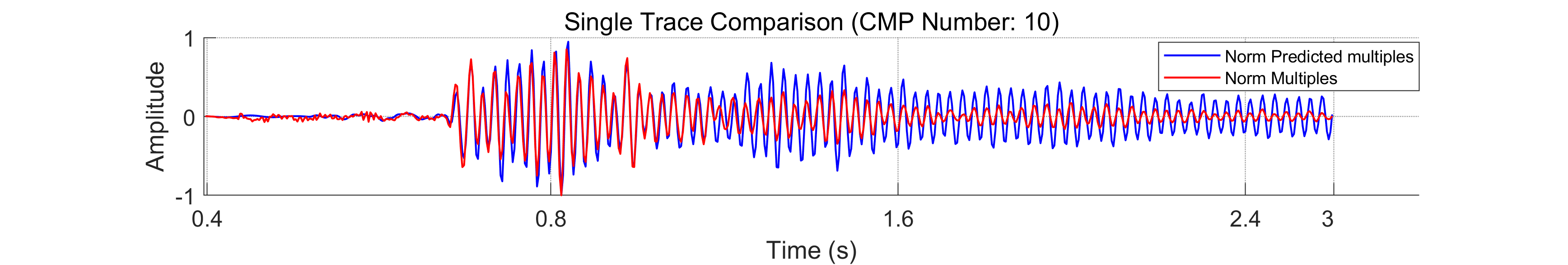}
    \caption{Waveform comparison of a single CMP trace (CMP 10) extracted from the stacking profiles in Figures \ref{fig12}c and \ref{fig12}d. The red curve represents the normalized multiples extracted by the proposed method, and the blue curve represents the normalized reference multiples predicted via the MDC operator. Both traces are normalized by their respective maximum absolute amplitudes.}
    \label{fig13}
\end{figure*}

To further quantify the suppression performance, nearest-offset stacking profiles are examined. Figure \ref{fig12}a shows the raw stacking profile, and Figure \ref{fig12}b presents the corresponding profile after multiple suppression by the proposed method. The extracted multiples, representing the difference between Figures \ref{fig12}a and \ref{fig12}b, are shown in Figure \ref{fig12}c. Figure \ref{fig12}d displays the reference multiples generated via the MDC operation. A comparison between Figures \ref{fig12}c and \ref{fig12}d demonstrates that the suppressed energy shares nearly identical spatial and temporal characteristics with the MDC-generated reference, confirming that the removed energy accurately corresponds to surface-related multiples rather than primaries.

For a trace-level waveform analysis, a single CMP trace (CMP 180) is extracted from both Figures \ref{fig12}c and \ref{fig12}d and compared in Figure \ref{fig13}, with both traces normalized by their respective maximum absolute amplitudes. The two traces exhibit a high degree of waveform consistency, with peaks and troughs showing excellent temporal alignment throughout the entire time range. This trace-level agreement provides final confirmation that the proposed method reliably captures and suppresses surface-related multiples in complex field environments, without damaging the underlying primary reflections.
\section{Discussion}\label{sec:discussion}
The proposed self-supervised learning (SSL) framework shares a fundamental principle with conventional surface-related multiple elimination (SRME): both derive predicted multiples directly from the recorded seismic data via convolution. However, the proposed approach offers two distinct advantages over conventional SRME workflows. First, it significantly enhances computational efficiency. Unlike iterative SRME-based approaches that require repeated multi-dimensional convolution (MDC) operations, our method performs only a single MDC operation to generate the initial multiple model, after which the neural network handles the suppression entirely. By avoiding repeated MDC operations, the computational burden is drastically reduced. Second, our method eliminates the dependency on source wavelet estimation. In conventional SRME, adaptive subtraction typically requires accurate wavelet information to reconcile the predicted and true multiples. As an SSL-based framework, our method requires neither wavelet estimation nor prior velocity information, making it inherently more robust for field data processing where such information is rarely available. These advantages are shared by both our method and the framework proposed by \cite{cheng2025self} (denoted by SSL-MDC), as our method builds upon their technical foundation, retaining the same network architecture and the self-supervised denoising strategy grounded in the Noisier2Noise principle.

Despite the shared foundations with SSL-MDC, the proposed framework introduces three improvements that directly address the practical limitations identified in Section \ref{sec:review}. Specifically, 
\begin{itemize}
    \item First, we replace manual scaling factor selection with an adaptive learning scheme. In SSL-MDC, the scaling factor range is determined empirically for each dataset, introducing subjectivity and requiring manual intervention to balance the amplitudes between the MDC-predicted and true multiples. We address this by treating $\alpha$ as a learnable parameter, jointly optimized with the network weights via gradient descent. Rather than being fixed or randomly sampled from a hand-crafted range, $\alpha$ evolves dynamically throughout training based on the gradient from the loss function, converging automatically to a value that best matches the amplitude relationship between $d_\text{m}$ and the true multiples for each specific dataset. This eliminates dataset-dependent empirical tuning and makes the framework directly applicable to unfamiliar data without prior inspection. 
    \item Second, we introduce a composite loss function with adaptive weighting to stabilize the optimization of the learnable $\alpha$. The loss consists of two $L_1$ terms: a regression term $L_\text{reg}$ that drives the network to learn the features of surface-related multiples, and a constraint term $L_\text{cons}$ that anchors $\alpha$ away from zero, preventing the degenerate solution in which the network learns to simply nullify $d_\text{m}$ rather than suppress multiples. To balance these two terms without manual hyperparameter search, we apply homoscedastic uncertainty-based weighting, which automatically modulates the contribution of each loss component throughout training and accommodates the varying loss scales that arise as $\alpha$ and the network weights co-evolve. 
    \item Third, the above two designs together streamline the training pipeline from two stages to one. In SSL-MDC, the iterative data refinement (IDR) stage is necessitated by the suboptimal suppression achieved during warm-up, which in turn stems from the imprecise manual selection of $\alpha$. By resolving the amplitude calibration issue at its root through the learnable $\alpha$, our warm-up stage alone produces suppression results of sufficient quality, rendering the computationally expensive IDR stage unnecessary. This reduces the total number of training epochs from 160 to 50 for the Otway model and from 80 to 50 for the layered model, substantially lowering the computational overhead while achieving comparable or superior suppression performance.
\end{itemize}
The experimental results on the Otway model and the Viking Graben field dataset (Figures \ref{fig5}, \ref{fig6}, and \ref{fig10}) further confirmed that the adaptive scaling strategy ensures the suppressed multiples are more accurately aligned with the MDC-based multiple predictions, while maintaining greater structural consistency with the reference subsurface images compared to SSL-MDC.

A natural question arises as to why the proposed method can dispense with the IDR stage and achieve satisfactory suppression using the warm-up stage alone. To understand this, it is instructive to examine the role that the IDR stage plays in SSL-MDC and why we can remove it in our framework. In SSL-MDC, the warm-up stage trains the network to map $d_\text{input} = d_\text{raw} + \alpha d_\text{m}$ back to $d_\text{raw}$. Although this follows the Noisier2Noise principle, the effectiveness of the learned mapping critically depends on how well the amplitude of $\alpha d_\text{m}$ matches that of the true multiples present in $d_\text{raw}$. When $\alpha$ is manually selected and suboptimal, the amplitude mismatch weakens the self-supervised training. As a result, the network cannot accurately characterize the amplitude properties of the multiples. The IDR stage is therefore introduced as a compensatory mechanism, iteratively replacing $d_\text{raw}$ with progressively cleaner network predictions to refine the pseudo-label quality over successive epochs. In this sense, the IDR stage is not an intrinsic requirement of the SSL paradigm, but rather a remedy for the imprecise amplitude calibration introduced by manual $\alpha$ selection. 

A deeper understanding of why the learnable $\alpha$ resolves this issue can be obtained by revisiting the theoretical conditions required by the Noisier2Noise principle. Specifically, it requires that the noise added to construct the input is statistically independent of the noise already present in the target. In the context of multiple suppression, the true multiples in $d_\text{raw}$ constitute the inherent noise in the target, while $\alpha d_\text{m}$ serves as the additional noise injected into the input. The condition is approximately satisfied since $\alpha d_\text{m}$ and the true multiples differ in amplitude and wavelet characteristics due to the absence of accurate source wavelets in MDC. Additionally, the gradual evolution of $\alpha$ during training introduces amplitude diversity into the training data, acting as an implicit regularizer that encourages the network to learn the morphological features of surface-related multiples rather than fitting a fixed amplitude ratio. As a result, the warm-up stage alone produces accurate suppression, rendering the IDR stage unnecessary.

While the proposed method demonstrates clear advantages over SSL-MDC in terms of automation, efficiency, and suppression quality, several limitations remain and point to directions for future research. On the one hand, the constraint loss $L_\text{cons}$ anchors $\alpha$ toward a fixed target value of 0.5, which is chosen empirically. Although this value has proven effective across the datasets examined in this study, its suitability for datasets with substantially different amplitude characteristics between $d_\text{m}$ and the true multiples has not been systematically investigated. A natural extension would be to replace this fixed target with a data-driven estimate, further reducing the remaining degree of manual specification. On the other hand, the current framework is formulated for two-dimensional seismic data. Extending the approach to three-dimensional acquisition, where surface-related multiples exhibit more complex azimuthal characteristics and the computational cost of MDC grows substantially, represents a significant and practically relevant direction for future work.
\section{Conclusions}\label{sec:conclusions}
We proposed an adaptive self-supervised learning (SSL) framework for surface-related multiple suppression. The framework generates predicted multiples via multi-dimensional convolution and trains a neural network to suppress them without requiring clean labeled data, wavelet estimation, or subsurface velocity models. Building upon a recently proposed SSL framework, the key innovation of our method lies in automatically determining the amplitude scaling between the predicted and true multiples through joint optimization with the network weights, eliminating the subjective manual selection required by the baseline method. To prevent this scaling from degenerating to a trivial solution, we designed a composite loss function with homoscedastic uncertainty-based adaptive weighting, which ensures stable and balanced optimization throughout training. Together, these designs enable the network to achieve robust multiple suppression in a single training stage, removing the need for the iterative data refinement stage required by the baseline method and substantially reducing the total training cost. Experiments on synthetic and field datasets demonstrated that the proposed method achieves comparable or superior suppression performance with significantly fewer training epochs, better preservation of primary reflections, and greater structural consistency in migrated images.
\section*{Acknowledgments}
This work was supported by the National Natural Science Foundation of China under Grant Nos. 42504133, 42130808, 42404140, 42574181. The author Shijun Cheng thanks King Abdullah University of Science and Technology for supporting this research.

\bibliographystyle{unsrtnat}
\bibliography{references}

\end{document}